\documentclass[aps,prd,12pt,nofootinbib]{revtex4-2}
\usepackage{bm}
\usepackage{amsmath}
\usepackage{cases}
\usepackage{ulem}
\usepackage{comment}
\usepackage{latexsym}
\usepackage{amsfonts}
\usepackage{graphicx}
\usepackage{mathrsfs}
\usepackage{CJKutf8}
\usepackage{subfigure}
\usepackage{float}
\usepackage{multirow}
\usepackage{orcidlink}
\usepackage{color}
\usepackage{booktabs}
\usepackage{hyperref}
\hypersetup{
	colorlinks=true,
	linkcolor=red,
	citecolor=blue,
}

\begin{document}

\begin{CJK*}{UTF8}{gbsn}

\title{More on difference between angular momentum and pseudo-angular momentum}
	\author{Qi Dai,${}^{*}$~Zi-Wei Chen\orcidlink{0000-0003-4752-0794},$^{*}$\textsuperscript{,}$^{\dagger}$ Bang-Hui Hua and Xiang-Song Chen$^{\dagger}$}
\renewcommand{\thefootnote}{\fnsymbol{footnote}}
\footnotetext[1]{These authors contributed equally to this work.}
\footnotetext[2]{Corresponding authors. E-mail: chenzw@hust.edu.cn (Zi-Wei Chen), cxs@hust.edu.cn (Xiang-Song Chen)}
\affiliation{School of Physics, Huazhong University of Science and Technology,
Wuhan 430074, China}

\begin{abstract}
	We extend the discussion on the difference between angular momentum and pseudo-angular momentum in field theory. We show that the often quoted expressions in [Phys.Rev.B \textbf{103}, L100409 (2021)] only apply to a non-linear system, and derive the correct rotation symmetry and the corresponding angular momentum for a linear elastic system governed by Navier-Cauchy equation. By mapping the concepts and methods for the elastic wave into electromagnetic theory, we argue that the renowned canonical and Benlinfante angular momentum of light are actually pseudo-angular momentum. Then, we derive the ``Newtonian" momentum $\int \text{d}^3 x\boldsymbol{E}$ and angular momentum $\int \text{d}^3 x (\boldsymbol{r}\times\boldsymbol{E})$ for a free electromagnetic wave, which are conserved quantities during propagation in vacuum.
\end{abstract}

\maketitle

\section{Angular Momentum in Classical Field Theory: From Electromagnet Wave to Elastic Wave}
What is a field's angular momentum? The answer can be easily found in textbooks on classical field theory\cite{soper2008classical,jackson2021classical}, or some pedagogical papers (see, e.g., \cite{LEADER2014163}). In Classical Electrodynamics, there are two renowned  expressions describing a free electromagnetic wave's angular momentum, those are 
\begin{numcases}{}
	\boldsymbol{J}_{\rm cano}=\int \text{d}^3 x \boldsymbol{r}\times E_i\boldsymbol{\nabla}A_i+\int \text{d}^3 x\boldsymbol{E}\times \boldsymbol{A}\equiv\boldsymbol{L}+\boldsymbol{S}\label{canoEM}\\\boldsymbol{J}_{\rm Bel}~~= \int \text{d}^3 x\boldsymbol{r}\times(\boldsymbol{E}\times\boldsymbol{B})\label{BelEM}.
\end{numcases}
Eq. (\ref{canoEM}) is the canonical angular momentum derived by Noether's theorem\cite{doi:10.1080/00411457108231446} through SO(3) symmetry, Eq. (\ref{BelEM}) is the Belinfante angular momentum which can be obtained from the canonical expression by the Belinfante symmetrization proceedure\cite{BELINFANTE1939887}, here and below latin indices run from 1 to 3 and Einstein summation rule is adopted. Aside from the gauge invariance problem, the canonical expression separates the spin angular momentum (SAM) and the orbital angular momentum (OAM) of electromagnetic field, while the Belinfante expression only gives the total angular momentum. \par
The SAM of light was experimentally detected early in 1936 by Beth\cite{PhysRev.50.115}, but it was not until the pioneering work of Allen \textit{et al}.\cite{PhysRevA.45.8185} in 1992 that the OAM of light could be physically realized by the helical wave-front structure with a phase singularity. Following similar but quick steps, recently, the SAM of elastic wave has been studied by Long \textit{et al}.\cite{long2018intrinsic}, then the OAM by Chaplain \textit{et al}.\cite{PhysRevLett.128.064301}. It is worth mentioning that elastic wave with inclined phase fronts can also carry an OAM just as light. In a subsequent paper\cite{PhysRevLett.129.204303}, Bliokh put forward the canonical total angular momentum of elastic wave (actually it should be regarded as pseudo-angular momentum, as Bliokh had already emphasized) by Noether's theorem with respect to spatial rotations:
\begin{equation} \label{pseudototalAMElastic}
	\textbf{\text{J}}=\int \text{d}^3 x\rho\boldsymbol{r}\times(-\dot{\xi}_i\boldsymbol{\nabla}\xi_i)+\int \text{d}^3 x\rho\boldsymbol{\xi}\times\dot{\boldsymbol{\xi}}\equiv\textbf{L}+\textbf{S},
\end{equation}
where $\rho$ is the density of the medium and  $\boldsymbol{\xi}(\boldsymbol{x},t)$ is the displacement field. \par
We can take a closer look at formula (\ref{canoEM}), (\ref{pseudototalAMElastic}) and notice that neither of the expressions behave like ``$\boldsymbol{r}\times(m\boldsymbol{v})$" in the sense of Newtonian mechanics. Even in formula (\ref{BelEM}), the Pointing vector $\boldsymbol{E}\times\boldsymbol{B}$, as Belinfante momentum density of light field, does not behave like some velocity. One may seriously ask, what is the quantity ``position$\times$velocity" for a classical field, since it is the original construction of angular momentum for a Newtonian particle?  Garanin and Chudnovsky\cite{PhysRevB.92.024421} first proposed the ``Newtonian" total angular momentum 
\begin{equation}  \label{NewtionianAM}
	\widetilde{\boldsymbol{\rm J}}=\int \text{d}^3 x\rho\boldsymbol{r}\times\dot{\boldsymbol{\xi}}+\int \text{d}^3 x\rho\boldsymbol{\xi}\times\dot{\boldsymbol{\xi}}\equiv\widetilde{\textbf{L}}+\widetilde{\textbf{S}},
\end{equation}
with a very clear picture (Fig. 1 in their paper) illustrating the motion that generates SAM and OAM in elastic solid. Nakane and Kohno (NK)\cite{PhysRevB.97.174403}  rederived the result using Noether's theorem with respect to a special rotation transformation:
\begin{equation} \label{Rota1}
	\boldsymbol{\xi}(\boldsymbol{x})\rightarrow\mathcal{R} \boldsymbol{\xi}(\boldsymbol{x})+(\mathcal{R}-\mathcal{I})\boldsymbol{x}, 
\end{equation}
where $\mathcal{R}$ is a constant SO(3) matrix and $\mathcal{I}$ is unit matrix. The above transformation can be compared with the conventional field rotation
\begin{equation}  \label{Rota2}
	\boldsymbol{\xi}(\boldsymbol{x})\rightarrow\mathcal{R}\boldsymbol{\xi}(\mathcal{R}^{-1}\boldsymbol{x}),
\end{equation}
which leads to the ``field-theory" expression (\ref{pseudototalAMElastic}). NK also gave an explanation on these two rotations: Eq. (\ref{Rota2}) is the rotation of wave patterns (or, field) with the equilibrium positions of medium particles fixed, and Eq. (\ref{Rota1}) corresponds to a rotation of entire medium including wave patterns (see, e.g., Fig. 1 in \cite{PhysRevB.103.L100409}). Then, Streib\cite{PhysRevB.103.L100409} addressed the issue of ``pseudo" physical quantity in detail, discussing not only the difference between  pseudo-momentum and momentum, but also pseudo-angular momentum (PAM) and angular momentum (AM).
Here we adopt the convention by defining conserved charges corresponding to field rotation as PAM, and charges corresponding to medium rotation as AM (which is called Newtonian angular momentum by NK). \par
The aim of this paper is to first point out that, although the above two rotation transformations are indeed symmetries for a complete elastic Lagrangian $\mathscr{L}$ (containing the anharmonic term in strain tensor), one of them fails to be a symmetry when the anharmonic part in strain tensor is disregarded; this is crucial because only the simplified Lagrangian $\mathcal{L}$ leads to the linear Navier-Cauchy equation. We therefore worked out the correct rotation symmetries for the simplified action $\int\text{d}^4 x \mathcal{L}$ and the corresponding AM. Moreover, we argue that the familiar expressions (\ref{canoEM}) and (\ref{BelEM}) for electromagnetic field are, in fact, PAM. Finally, inspired by elastic AM, the momentum (which is actually the Newtonian momentum) and AM of a free electromagnetic wave are also derived. \par
This paper is organized as follows. In section 2, we discuss PAM and AM of two different elastic systems governed respectively by Lagrangian $\mathscr{L}$ and $\mathcal{L}$. In section 3, we derive the momentum and AM of a free electromagnetic field in vacuum. A summary and discussion are given in section 4.

	\section{Angular Momentum Versus Pseudo-angular Momentum of Elastic Wave}
Consider an elastic system which is governed by Lagrangian
\begin{equation}
	\mathscr{L}=\frac{1}{2}\rho\dot{\xi_{i}}\cdot\dot{\xi_{i}}-\mu u_{ij}\cdot u_{ij}-\frac{1}{2}\lambda u_{ii}\cdot u_{jj},
\end{equation}
where 
\begin{equation}
	u_{ij}=\dfrac{1}{2}\left(\partial_{i}\xi_{j}+\partial_{j}\xi_{i}+\partial_{i}\xi_{l}\partial_{j}\xi_{l}\right)
\end{equation}
is the strain tensor, $\lambda,\mu$ are Lam\'{e}'s first and second parameters\cite{landau1986theory}. The Euler-Lagrange equation for displacement field is  
\begin{equation}
	\dfrac{\partial \mathscr{L}}{ \partial\xi_{i}}-\partial_{\mu}\frac{\partial\mathscr{L}}{\partial\left(\partial_{\mu}\xi_i \right) }=0,
\end{equation}
here and below Greek indices run from 0 to 3. Plug in $\mathscr{L}$, we have equation of motion as
\begin{equation} \label{nonlinearEOM}
	\partial_{j}\left[\lambda u_{ll}\left(\delta_{ij}+\partial_{j}\xi_{i}\right)+2\mu\left(u_{ij}+u_{jm}\partial_{m}\xi_{i}\right) \right]=\rho\ddot{\xi_i}.
\end{equation}
Note that Eq. (\ref{nonlinearEOM}) is a non-linear dynamic equation distinct from Navier-Cauchy equation. \par
As discussed in paper \cite{PhysRevB.97.174403}, this elastic system is invariant under two kinds of rotations, the first one is rotation of  field pattern only (cf. Eq. (\ref{Rota2})) , which results in elastic PAM (\ref{pseudototalAMElastic}). The second one is rotation of entire medium (cf. Eq. (\ref{Rota1})), which results in elastic AM (\ref{NewtionianAM}). \par
If we regard the displacement field $\xi_i$ as infinitesimal quantities and ignore higher order terms in $u_{ij}$, i.e.
\begin{equation}
	u_{ij}=\dfrac{1}{2}\left(\partial_{i}\xi_{j}+\partial_{j}\xi_{i}\right),
\end{equation}
the Lagrangian becomes
\begin{equation}
	\mathcal{L}=\frac{1}{2}\rho\dot{\xi_{i}}\cdot\dot{\xi_{i}}-\frac{1}{2}\lambda\partial_i\xi_i\cdot\partial_j\xi_j-\frac{1}{4}\mu%
	\left(\partial_{j}\xi_{i}+\partial_{i}\xi_{j}\right)^2.\label{Lagrangian}
\end{equation}
The corresponding equation of motion of $\mathcal{L}$ is the well-known Navier-Cauchy equation:
\begin{equation}
	\mu \partial_j\partial_j \xi_{i}+(\lambda+\mu)\partial_i\partial_j\xi_j=\rho\ddot{\xi}_i.
\end{equation}
\par
Rotation (\ref{Rota2}) is still a symmetry of such linear elastic system, and it gives the same conserved charge as Eq. (\ref{pseudototalAMElastic}). However, rotation (\ref{Rota1}) fails to be a symmetry because we have ignored the higher order terms in $u_{ij}$. The proper rotation symmetry can be verified as 
\begin{equation} \label{lrotation1}
	\boldsymbol{\xi}(\boldsymbol{x})\rightarrow\boldsymbol{\xi}(\boldsymbol{x})+(\mathcal{R}-\mathcal{I})\boldsymbol{x},
\end{equation}
and it leads to AM
\begin{equation}  \label{NewtionianAMlinear}
	\widetilde{\boldsymbol{\rm J}}_{\rm linear}=\int \text{d}^3 x\rho\boldsymbol{r}\times\dot{\boldsymbol{\xi}}.
\end{equation}
Eq. (\ref{NewtionianAMlinear}) indicates that, the orbital-like expression $\int\text{d}^3 x (\rho\boldsymbol{r}\times\dot{\boldsymbol{\xi}})$ is conserved in this linear system. \par
We list the above results in Table \ref{table:1} for comparison. 
\begin{table}[H]
	\begin{center}
		\caption{\small Comparison between non-linear and linear elastic system. For both systems, the field rotation symmetry (Rotation 1), pseudo OAM and pseudo SAM are the same. Specially, in linear system, one cannot clearly separate OAM and SAM for Rataion 2, leaving a total AM expression.}
		\label{table:1}
		\renewcommand{\arraystretch}{0.81}	
		\scalebox{0.95}{ 
			\begin{tabular}{c|c|c}
				\hline\hline
				System&Non-linear&Linear\\ \hline
				Lagrangian&$\mathscr{L}=\frac{1}{2}\rho\dot{\xi_{i}}\cdot\dot{\xi_{i}}-\mu u_{ij}\cdot u_{ij}-\frac{1}{2}\lambda u_{ii}\cdot u_{jj}$ & $\mathcal{L}=\frac{1}{2}\rho\dot{\xi_{i}}\cdot\dot{\xi_{i}}-\mu
				u_{ij}\cdot u_{ij}-\frac{1}{2}\lambda u_{ii}\cdot u_{jj}$\\
				Strain tensor& $u_{ij}=\frac{1}{2}(\partial_{i}\xi_{j}+\partial_{j}\xi_{i}+\partial_{i}\xi_{l}\partial_{j}\xi_{l})$ & $u_{ij}=\frac{1}{2}(\partial_{i}\xi_{j}+\partial_{j}\xi_{i})$  \\
				\hline
				Rotation 1 &$\boldsymbol{\xi}(\boldsymbol{x})\rightarrow\mathcal{R}\boldsymbol{\xi}(\mathcal{R}^{-1}\boldsymbol{x})$&$\boldsymbol{\xi}(\boldsymbol{x})\rightarrow\mathcal{R}\boldsymbol{\xi}(\mathcal{R}^{-1}\boldsymbol{x})$
				\\
				
				Pseudo OAM density& $-\rho\boldsymbol{r}\times(\dot{\xi}_i\boldsymbol{\nabla}\xi_i)$&$-\rho\boldsymbol{r}\times(\dot{\xi}_i\boldsymbol{\nabla}\xi_i)$ \\
				Pseudo SAM density&   $\rho\boldsymbol{\xi}\times\dot{\boldsymbol{\xi}}$ &$\rho\boldsymbol{\xi}\times\dot{\boldsymbol{\xi}}$ \\
				\hline
				Rotation 2 & $		\boldsymbol{\xi}(\boldsymbol{x})\rightarrow\mathcal{R} \boldsymbol{\xi}(\boldsymbol{x})+(\mathcal{R}-\mathcal{I})\boldsymbol{x}$ & $\boldsymbol{\xi}(\boldsymbol{x})\rightarrow\boldsymbol{\xi}(\boldsymbol{x})+(\mathcal{R}-\mathcal{I})\boldsymbol{x}$ \\
				
				OAM density& $\rho\boldsymbol{r}\times\dot{\boldsymbol{\xi}}$ & -- \\
				SAM density&  $\rho\boldsymbol{\xi}\times\dot{\boldsymbol{\xi}}$  &--\\ Total AM density& $\rho\boldsymbol{r}\times\dot{\boldsymbol{\xi}}+\rho\boldsymbol{\xi}\times\dot{\boldsymbol{\xi}}$&$\rho\boldsymbol{r}\times\dot{\boldsymbol{\xi}}$\\
				\hline\hline
			\end{tabular}
		}
	\end{center}
\end{table}

\section{Electromagnetic Momentum and Angular Momentum Revisited: Hints From Elastic Wave}
Elastic wave also carries momentum and pseudo-momentum, let us take the linear system for example, the pseudo-momentum corresponds to symmetric transformation
\begin{equation}
	\boldsymbol{\xi}(\boldsymbol{x}) \rightarrow\boldsymbol{\xi}(\boldsymbol{x}+\boldsymbol{a}),
\end{equation}
where $\boldsymbol{a}$ is an infinitesimal constant vector; the momentum corresponds to 
\begin{equation}
	\boldsymbol{\xi}(\boldsymbol{x}) \rightarrow\boldsymbol{\xi}(\boldsymbol{x})+\boldsymbol{a}.
\end{equation}
\par
Discussion on physical quantities and pseudo quantities in elastic theory is very illuminating. Thus, we are now to map the concepts and methods into electromagnetic theory. Consider a free electromagnetic wave in vacuum described by
\begin{equation}
	\mathscr{L}_{\text{EM}}=-\frac{1}{4}\mathcal{F}_{\mu\nu}\mathcal{F}^{\mu\nu},
\end{equation}
where $\mathcal{F}_{\mu\nu}=\partial_\mu A_\nu-\partial_\nu A_\mu$. We find that, a  ``rotation"  similar to (\ref{lrotation1}),
\begin{equation} \label{spectialrotaEM}
	\boldsymbol{A}(\boldsymbol{x})\rightarrow\boldsymbol{A}(\boldsymbol{x})+(\mathcal{R}-\mathcal{I})\boldsymbol{x},
\end{equation}
keeps the action invariant. The AM of electromagnetic wave then reads 
\begin{equation}
	\boldsymbol{\widetilde{J}}=\int \text{d}^3 x\boldsymbol{r}\times\partial_{t}\boldsymbol{A},
\end{equation}
where $\partial_{t}\boldsymbol{A}$ is similar to $\dot{\boldsymbol{\xi}}$ in Eq. (\ref{NewtionianAMlinear}). The above expression can also be written as 
\begin{equation}
	\boldsymbol{\widetilde{J}}=\int \text{d}^3 x\boldsymbol{r}\times\boldsymbol{E}
\end{equation}
in Coulomb gauge ($\partial_t\boldsymbol{A}=\boldsymbol{E}$ is used). Recall that $\boldsymbol{E}$ is the conjugate momentum of $\boldsymbol{A}$ in Hamiltonian mechanics, then $\boldsymbol{\widetilde{J}}$ is of the form ``position$\times$momentum". Here, one must note: 1) we are considering electromagnetic wave in vacuum, not any medium is rotating, but the rotation transformation can still be applied mathematically; 2) the natural units $\hbar=c=1$, and electrostatic constant $k=1$ are used throughout this paper, thus we have ignored the constants in (\ref{spectialrotaEM}) for matching dimension. \par
For translation symmetry corresponding to the ``medium"
\begin{equation}
	\boldsymbol{A}(\boldsymbol{x})\rightarrow\boldsymbol{A}(\boldsymbol{x})+\boldsymbol{a},
\end{equation} 
one can promptly calculate the conserved momentum as
\begin{equation}
	\boldsymbol{\widetilde{P}}=\int \text{d}^3 x \partial_t\boldsymbol{A}.
\end{equation}
Similarly, in Coulomb gauge we have
\begin{equation}
	\boldsymbol{\widetilde{P}}=\int \text{d}^3 x\boldsymbol{E}.
\end{equation}
\par
One should note that $\widetilde{\boldsymbol{P}}$ is merely a trivial momentum for a free electromagnetic wave packet since 
$
\int \text{d}^3 x \boldsymbol{E} = \int \text{d}^3 x \partial_i(\boldsymbol{x}E_i)=\boldsymbol{0}
$.

We list our results in Tabel \ref{table:2} for comparison.
\begin{table}[H]
	\begin{center}
		\caption{\small Comparison of elastic and electromagnetic systems}
		\label{table:2}
		\renewcommand{\arraystretch}{0.81}	
		\scalebox{0.95}{ 
			\begin{tabular}{c|c|c}
				\hline\hline
				System& Elastic & Electromagnetic\\ \hline
				Translation 1 &  $\boldsymbol{\xi}(\boldsymbol{x}) \rightarrow\boldsymbol{\xi}(\boldsymbol{x}+\boldsymbol{a})$&$\boldsymbol{A}(\boldsymbol{x}) \rightarrow\boldsymbol{A}(\boldsymbol{x}+\boldsymbol{a})$ \\
				
				Pseudo-Momentum density& $ -\rho\dot{\xi}_i\boldsymbol{\nabla}\xi_i$ & $E_i\boldsymbol{\nabla}A_i$ \\\hline
				Translation 2 &  $\boldsymbol{\xi}(\boldsymbol{x}) \rightarrow\boldsymbol{\xi}(\boldsymbol{x})+\boldsymbol{a}$&$\boldsymbol{A}(\boldsymbol{x})\rightarrow\boldsymbol{A}(\boldsymbol{x})+\boldsymbol{a}$ \\
				
				Momentum density& $\rho \dot{\boldsymbol{\xi}}$ & $\boldsymbol{E}$ \\\hline\hline
				Rotation 1 &{$\boldsymbol{\xi}(\boldsymbol{x})\rightarrow\mathcal{R}\boldsymbol{\xi}(\mathcal{R}^{-1}\boldsymbol{x})$}&$\boldsymbol{A}(\boldsymbol{x})\rightarrow\mathcal{R}\boldsymbol{A}(\mathcal{R}^{-1}\boldsymbol{x})$
				\\
				
				Pseudo OAM density& {$\rho\boldsymbol{r}\times(-\dot{\xi}_i\boldsymbol{\nabla}\xi_i)$ }& $\boldsymbol{r}\times E_i\boldsymbol{\nabla}A_i$\\
				Pseudo SAM density&   {$\rho\boldsymbol{\xi}\times\dot{\boldsymbol{\xi}}$} & $\boldsymbol{E}\times\boldsymbol{A}$\\
				\hline
				Rotation 2&  $\boldsymbol{\xi}(\boldsymbol{x})\rightarrow\boldsymbol{\xi}(\boldsymbol{x})+(\mathcal{R}-\mathcal{I})\boldsymbol{x}$&$\boldsymbol{A}(\boldsymbol{x})\rightarrow\boldsymbol{A}(\boldsymbol{x})+(\mathcal{R}-\mathcal{I})\boldsymbol{x}$ \\
				
				AM density& $\rho\boldsymbol{r}\times\dot{\boldsymbol{\xi}}$ & $\boldsymbol{r}\times\boldsymbol{E}$ \\
				
				\hline\hline
			\end{tabular}
		}
	\end{center}
\end{table}

\section{Summary and Discussion}
We summarize our observations in this paper:\par
($\text{i}$) For a classical field $\boldsymbol{f}(\boldsymbol{x},t)$, the commonly used angular momentum derived from symmetry of field pattern rotation $\mathcal{R}\boldsymbol{f}(\mathcal{R}^{-1}\boldsymbol{x})$, should be regarded as PAM; momentum that derived from the symmetry of usual field translation  $\boldsymbol{f}(\boldsymbol{x}+\boldsymbol{a})$ should be regarded as pseudo-momentum. It means that the renowed canonical angular momentum of light (or even the Belinfante angular momentum) is, in fact, PAM.\par
($\text{ii}$) At least for a free electromagnetic field and elastic field, we have worked out the  Newtonian angular momentum. The subtle thing is, a linear elastic system $\mathcal{L}$ and a non-linear one $\mathscr{L}$, do not share the same rotation symmetries, and we find out a new rotation symmetry $\boldsymbol{\xi}(\boldsymbol{x})\rightarrow\boldsymbol{\xi}(\boldsymbol{x})+(\mathcal{R}-\mathcal{I})\boldsymbol{x}$ for $\mathcal{L}$. Remarkably, such rotation symmetry can also be found in free electromagnetic systems as $	\boldsymbol{A}(\boldsymbol{x})\rightarrow\boldsymbol{A}(\boldsymbol{x})+(\mathcal{R}-\mathcal{I})\boldsymbol{x}$. As a result, the quantity $\int \text{d}^3 x (\boldsymbol{r}\times\boldsymbol{E})$ conserves for a free propagating electromagnetic wave packet. \par
($\text{iii}$) The momentum of free electromagnetic field and elastic field are also addressed, resulting in two velocity-like expressions, $\rho\dot{\boldsymbol{\xi}}$ for elastic wave and $\partial_t\boldsymbol{A}$ (or $\boldsymbol{E}$) for electromagnetic wave.\par
Finally, we suggest that it might be interesting to further consider the similar issue for spin-1/2 fields: whether the difference between physical quantities and pseudo quantities still exists, and are there any useful conserved charges which were not noticed before?

\end{CJK*}

\begin{acknowledgments}
	This work is supported by the China NSF via Grants No. 11535005.
\end{acknowledgments}

%
\end{document}